\documentclass[11pt]{article}
\usepackage{jcappub}
\usepackage{lineno,hyperref}
\modulolinenumbers[5]
\usepackage{cancel}
\usepackage{fancyvrb}
\usepackage{upquote}
\usepackage[utf8]{inputenc} 
\usepackage{geometry}
\usepackage{array}
\usepackage{multirow}
\newcolumntype{P}[1]{>{\centering\arraybackslash}p{#1}}
\newcolumntype{M}[1]{>{\centering\arraybackslash}m{#1}}

\usepackage{xcolor}
\newcommand{\be}{\begin{equation}}
\newcommand{\ee}{\end{equation}}
\newcommand{\een}{\end{subequations}}
\newcommand{\ben}{\begin{subequations}}
\newcommand{\beq}{\begin{eqalignno}}
\newcommand{\eeq}{\end{eqalignno}}

\newcommand{\lsim}{\mathrel{\mathop{\kern 0pt \rlap
      {\raise.2ex\hbox{$<$}}}\lower.9ex\hbox{\kern-.190em $ \sim$}}}
\newcommand{\gsim}{\mathrel{\mathop{\kern 0pt
      \rlap{\raise.2ex\hbox{$>$}}}\lower.9ex\hbox{\kern-.190em $\sim$}}}

\newcommand{\CO}{\mathcal{O}}

\newcommand{\erf}{\mbox{erf}}
\newcommand{\op}{{\cal O}}

\newcommand{\C}{{\cal C}}
\newcommand{\Q}{{\cal Q}}

\newcommand{\todo}[1]{{\color{red} \ifmmode\else[todo]\fi #1}}

\usepackage{fancyhdr}
\title{A DAMA/Libra--phase2 analysis in terms of WIMP–quark and
  WIMP–gluon effective interactions up to dimension seven}

\author{Sunghyun Kang,}
\author{Stefano Scopel,}
\author{Gaurav Tomar,}
\affiliation{Department of Physics, Sogang University, 
Seoul, Korea, 121-742}

\emailAdd{francis735@naver.com}
\emailAdd{scopel@sogang.ac.kr}
\emailAdd{tomar@sogang.ac.kr}

\abstract{ We analyze the DAMA/Libra--phase2 modulation result using a
  basis of 16 effective operators describing the WIMP interaction with
  photons, gluons and quarks up to mass dimension seven. For each
  operator we fix the effective theory at the scale of 2 GeV and
  parametrize WIMP--quark interactions in terms of two independent
  couplings for up--type and down--type quarks. We discuss the
  connection with the non--relativistic limit of the effective theory
  in terms of operators invariant by Galilean transformations, and the
  impact of the ensuing momentum--dependent Wilson coefficients due to
  long--range interactions or light meson poles. Most relativistic
  operators yield a good fit of the DAMA modulation effect, although
  for parameters excluded by the constraints from XENON1T and PICO60.}
\begin{document}
\maketitle

\section{Introduction}
\label{sec:intro}

The Large Hadron Collider (LHC) did not find so far any evidence of
physics beyond the Standard Model, excluding the most straightforward
realizations of Supersymmetry or Large Extra Dimensions that provided
the most popular Dark Matter (DM) candidates. As a consequence,
model-independent approaches have become increasingly popular to
interpret DM search experiments~\cite{chang_momentum_dependence_2010,
  dobrescu_nreft, fan_2010, hill_solon_nreft, peter_nreft,
  cirelli_tools_2013, effective_wimps_2014, catena_nreft,
  catena_directionality_nreft_2015, Catena_Gondolo_global_fits,
  cerdeno_nreft, Catena_Gondolo_global_limits, nreft_bayesian,
  xenon100_nreft,
  cresst_nreft,matching_solon1,matching_solon2,chiral_eft,hoferichter_si,bishara_2017}.
In particular, the DAMA modulation excess may serve as a benchmark to
develop and test such approaches~\cite{systematic_inelastic_sogang,
  eft_spin, dama_2018_sogang, dama_inelastic_eft_sogang}. It has
reached a more than 12$\sigma$ statistical significance and for it no
explanation alternative to a WIMP signal has been found so far, while
experiments trying to replicate it using the same sodium iodide target
material have only now started to reach the required
sensitivity~\cite{cosine_modulation_2019, cosine100_dama,
  anais_modulation_2019}. Nevertheless, in spite of its potential
relevance, a completely model--independent assessment of such result
against the null outcomes of other experiments is still not available
after almost 20 years of its appearance.

In order to address this aspect the DAMA result has been analyzed
parametrizing the WIMP--nucleus interaction in terms of the
non--relativistic operators allowed by Galilean
invariance~\cite{haxton1, haxton2} including all interferences
~\cite{Catena_dama, dama_inelastic_eft_sogang}, finding that, in the
case of a Maxwellian WIMP velocity distribution, the tension between
the WIMP interpretation of the modulation effect and the constraints
from other null results persists. The same result was found assuming
dominance of one non--relativistic operator at a
time~\cite{dama_2018_sogang}.  In all such analyses, however, no
explicit momentum dependence of the Wilson coefficients of the Non
Relativistic (NR) Effective Field Theory (EFT) was assumed.  If, in
alternative, the WIMP-nucleus interaction is parameterized in terms of
a relativistic WIMP--quark, WIMP-photon or WIMP--gluon effective
theory~\cite{tait_2010, buckley_2013, desimone_2016, belyaev_reft,
  bishara_2017, reft_dim7} (i) only a few of the non--relativistic
operators allowed by Galilean invariance actually appear in the
non--relativistic limit ; (ii) usually, more than one operator appears
in the non--relativistic limit, including interference terms; (iii)
the couplings of the non--relativistic theory may acquire an explicit
momentum dependence due to a long--range component of the relativistic
coupling or because of light meson poles. In this case an assessment
on the DAMA fit, including the latest DAMA/Libra phase2 low--threshold
analysis, has not been provided yet. To fill this gap in the present
paper we wish to extend the analysis of Ref.~\cite{dama_2018_sogang}
with an assessment of the DAMA/Libra--phase2 modulation result using a
basis of 16 effective operators describing the WIMP effective
interaction with photons, gluons and quarks up to mass dimension
seven:

\begin{equation}\label{eq:lightDM:Lnf5}
{\cal L}_\chi=\sum_q \sum_{a,d}
\C_{a,q}^{(d)} {\cal Q}_{a,q}^{(d)}+\sum_{b,d}
\C_{b}^{(d)} {\cal Q}_{b}^{(d)}, 
\end{equation}
\noindent

In particular, we assume the same set of operators discussed
in~\cite{bishara_2017} and analyzed
in~\cite{sogang_scaling_law_rel}. Analyses on similar sets of
relativistic DM effective operators can also be found
in~\cite{tait_2010, buckley_2013, desimone_2016, belyaev_reft}.

The paper is organized as follows: in Section~\ref{sec:eft} we list
the ${\cal Q}_{a,q}^{(d)}$, ${\cal Q}_{b}^{(d)}$ effective operators
and outline how we calculate the corresponding DM Direct Detection
(DD) expected rates; Section~\ref{sec:analysis} contains our
quantitative results; we draw our conclusions in
Section~\ref{sec:conclusions}.

\section{WIMP rates in effective models}
\label{sec:eft}

We consider the two dimension-five operators:
\begin{equation}
\label{eq:dim5}
{\cal Q}_{1}^{(5)} = \frac{e}{8 \pi^2} (\bar \chi \sigma^{\mu\nu}\chi)
 F_{\mu\nu} \,, \qquad {\cal Q}_2^{(5)} = \frac{e }{8 \pi^2} (\bar
\chi \sigma^{\mu\nu} i\gamma_5 \chi) F_{\mu\nu} \,,
\end{equation}
where $F_{\mu\nu}$ is the electromagnetic field strength tensor and
$\chi$ is the DM field, assumed here to be a Dirac particle. Such
operators correspond, respectively, to magnetic--dipole and
electric--dipole DM and imply a long--range
interaction~\cite{delnobile_2018} \footnote{The anapole coupling
  $(\bar \chi \gamma^{\mu}\gamma_5\chi) \partial^{\nu}F_{\mu\nu}$
  leads instead to an effective contact interaction. The DAMA result
  was already analyzed in the cases of magnetic dipole and anapole DM
  in~\cite{anapole_2014}, while a recent analysis of anapole DM is
  also provided in~\cite{sogang_gondolo_anapole}.}.  The dimension-six
operators are
\begin{eqnarray}
{\cal Q}_{1,q}^{(6)} & =& (\bar \chi \gamma_\mu \chi) (\bar q \gamma^\mu q)\,,
 {\cal Q}_{2,q}^{(6)} = (\bar \chi\gamma_\mu\gamma_5 \chi)(\bar q \gamma^\mu q)\,, \nonumber
  \\ 
{\cal Q}_{3,q}^{(6)} & =& (\bar \chi \gamma_\mu \chi)(\bar q \gamma^\mu \gamma_5 q)\,,
   {\cal Q}_{4,q}^{(6)} = (\bar
\chi\gamma_\mu\gamma_5 \chi)(\bar q \gamma^\mu \gamma_5 q)\,,\label{eq:dim6}
\end{eqnarray}
and we also include the following dimension-seven operators:
namely: 
\begin{eqnarray}
{\cal Q}_1^{(7)} & =& \frac{\alpha_s}{12\pi} (\bar \chi \chi)
 G^{a\mu\nu}G_{\mu\nu}^a\,, 
  {\cal Q}_2^{(7)} = \frac{\alpha_s}{12\pi} (\bar \chi i\gamma_5 \chi) G^{a\mu\nu}G_{\mu\nu}^a\,,\nonumber
 \\
{\cal Q}_3^{(7)} & =& \frac{\alpha_s}{8\pi} (\bar \chi \chi) G^{a\mu\nu}\widetilde
 G_{\mu\nu}^a\,, 
 {\cal Q}_4^{(7)} = \frac{\alpha_s}{8\pi}
(\bar \chi i \gamma_5 \chi) G^{a\mu\nu}\widetilde G_{\mu\nu}^a \,, \nonumber
\\
{\cal Q}_{5,q}^{(7)} & =& m_q (\bar \chi \chi)( \bar q q)\,, 
{\cal
  Q}_{6,q}^{(7)} = m_q (\bar \chi i \gamma_5 \chi)( \bar q q)\,,\nonumber
  \\
{\cal Q}_{7,q}^{(7)} &=& m_q (\bar \chi \chi) (\bar q i \gamma_5 q)\,, 
{\cal Q}_{8,q}^{(7)}  = m_q (\bar \chi i \gamma_5 \chi)(\bar q i \gamma_5
q)\,, \nonumber  
 \\
{\cal Q}_{9,q}^{(7)} & =& m_q (\bar \chi \sigma^{\mu\nu} \chi) (\bar q \sigma_{\mu\nu} q)\,, 
{\cal Q}_{10,q}^{(7)}  = m_q (\bar \chi  i \sigma^{\mu\nu} \gamma_5 \chi)(\bar q \sigma_{\mu\nu}
q)\,. \label{eq:dim7} 
\end{eqnarray}
\noindent In the equations above $q=u,d,s$ denote the light quarks,
$G_{\mu\nu}^a$ is the QCD field strength tensor, while $\widetilde
G_{\mu\nu} = \frac{1}{2}\varepsilon_{\mu\nu\rho\sigma} G^{\rho\sigma}$
is its dual, and $a=1,\dots,8$ are the adjoint color indices. In the
following we will also assume that all the operators listed in
Eqs.(\ref{eq:dim5})--(\ref{eq:dim7}) conserve flavor.

In the following Section we will show our results parametrizing each
coupling $\C^{(d)}$ = $\C^{(d)}_{a,q}$, $\C^{(d)}_{b}$ in
Eq.~(\ref{eq:lightDM:Lnf5}) with the effective scale
$\tilde{\Lambda}$:

\begin{equation}
  \C^{(d)}\equiv \frac{1}{\tilde{\Lambda}^{d-4}}.
  \label{eq:lambda}
\end{equation}

\noindent In particular, for the interactions with quarks we factorize
the coupling $\C^{(d)}_{up}$=1/$\tilde{\Lambda}^{d-4}$ with the up
quark and assume a common value $\C^{(d)}_{down}$=$\C^{(d)}_{strange}$
for two remaining couplings with the down and strange quarks,
introducing the additional parameter:

\begin{equation}
r=\C^{(d)}_{down}/\C^{(d)}_{up}.
\label{eq:r}
  \end{equation}

The non--relativistic limit of each of the operators of
Eqs.~(\ref{eq:dim5}--\ref{eq:dim7}) yields a linear combination of the
operators that parameterize the most general effective Hamiltonian for
the WIMP--nucleus interaction that complies with Galilean symmetry,
containing at most 15 terms in the case of a spin--1/2
particle~\cite{haxton1, haxton2}:

\begin{eqnarray}
{\bf\mathcal{H}}({\bf{r}})&=& \sum_{\tau=0,1} \sum_{j=1}^{15}
c_j^{\tau} \mathcal{O}_{j}({\bf{r}}) \, t^{\tau}.
\label{eq:H}
\end{eqnarray}

\begin{table}[]
\begin{center}
\begin{tabular}{|l|l|l|l|l|l|l|l|}
\hline\hline
\multicolumn{4}{l|}{\multirow{7}{*}{}} & \multicolumn{4}{l}{\multirow{7}{*}{}} \\
\multicolumn{4}{l|}{$ \CO_1 = 1_\chi 1_N$} \\
\multicolumn{4}{l|}{$\CO_2 = (v^\perp)^2$} & \multicolumn{4}{l}{$\CO_9 = i \vec{S}_\chi \cdot (\vec{S}_N \times {\vec{q} \over m_N})$}\\
\multicolumn{4}{l|}{$\CO_3 = i \vec{S}_N \cdot ({\vec{q} \over m_N} \times \vec{v}^\perp)$} & \multicolumn{4}{l}{$\CO_{10} = i \vec{S}_N \cdot {\vec{q} \over m_N}$} \\
\multicolumn{4}{l|}{$\CO_4 = \vec{S}_\chi \cdot \vec{S}_N$} & \multicolumn{4}{l}{$\CO_{11} = i \vec{S}_\chi \cdot {\vec{q} \over m_N}$} \\
\multicolumn{4}{l|}{$\CO_5 = i \vec{S}_\chi \cdot ({\vec{q} \over m_N} \times \vec{v}^\perp)$} & \multicolumn{4}{l}{$\CO_{12} = \vec{S}_\chi \cdot (\vec{S}_N \times \vec{v}^\perp)$} \\
\multicolumn{4}{l|}{$\CO_6=
  (\vec{S}_\chi \cdot {\vec{q} \over m_N}) (\vec{S}_N \cdot {\vec{q} \over m_N})$} & \multicolumn{4}{l}{$\CO_{13} =i (\vec{S}_\chi \cdot \vec{v}^\perp  ) (  \vec{S}_N \cdot {\vec{q} \over m_N})$} \\
\multicolumn{4}{l|}{$\CO_7 = \vec{S}_N \cdot \vec{v}^\perp$} & \multicolumn{4}{l}{$\CO_{14} = i ( \vec{S}_\chi \cdot {\vec{q} \over m_N})(  \vec{S}_N \cdot \vec{v}^\perp )$} \\
\multicolumn{4}{l|}{$\CO_8 = \vec{S}_\chi \cdot \vec{v}^\perp$} & \multicolumn{4}{l}{$\CO_{15} = - ( \vec{S}_\chi \cdot {\vec{q} \over m_N}) ((\vec{S}_N \times \vec{v}^\perp) \cdot {\vec{q} \over m_N})$} \\ \hline
\end{tabular}
\caption{Non-relativistic Galilean invariant operators for dark matter with spin $1/2$.}
\label{table:operators}
\end{center}
\end{table}

\noindent In the equation above the $\mathcal{O}_{j}$ operators are
listed in Table~\ref{table:operators}\cite{haxton2} and $t^0=1$,
$t^1=\tau_3$ denote the $2\times2$ identity and third Pauli matrix in
isospin space, respectively, and the isoscalar and isovector coupling
constants $c^0_j$ and $c^{1}_j$, are related to those to protons and
neutrons $c^{p}_j$ and $c^{n}_j$ by $c^{p}_j=(c^{0}_j+c^{1}_j)/2$ and
$c^{n}_j=(c^{0}_j-c^{1}_j)/2$. In general, for a given relativistic
operator $\Q^{(d)}$ the ensuing coefficients $c_j^{\tau}$ may depend
on the WIMP mass and/or the exchanged momentum $q\equiv |\vec{q}|$.
In Table~\ref{table:operator_correspondence} we summarize such
correspondence. We choose for convenience to adopt the same
notation used in Ref.~\cite{bishara_2017}, and the explicit expression
of the low--energy coefficients can be found in the Appendix A of the
same paper, or in~\cite{cirelli_tools_2013}. In particular, for their
numerical calculation we will use the output of the code
DirectDM~\cite{directdm}.

\begin{table}[t]
\begin{center}
\begin{tabular}{l}
  \hline\hline
$\Q_{1}^{(5)}\to -\frac{\alpha}{2\pi} F_1^{N}\Big(
\frac{1}{m_\chi}\op_1^N-4 \frac{m_N}{\vec q\,^2}
\op_5^N\Big)-\frac{2\alpha}{\pi
}\frac{\mu_N}{m_N}\Big(\op_4^N-\frac{m_N^2}{\vec q\,^2}
\op_6^N\Big)+{\mathcal O}(q^2)$\\
$\Q_{2}^{(5)}\to\frac{2\alpha}{\pi}\frac{m_N}{\vec q\,^2} F_1^N \op_{11}^N+{\mathcal O}(q^2)$\\
$\Q_{1,q}^{(6)}\to F_1^{q/N}\op_1^N+{\mathcal O}(q^2)$\\
$\Q_{2,q}^{(6)}\to 2 F_1^{q/N}\op_8^N+ 2\big(F_1^{q/N}+F_2^{q/N}\big) \op_9^N+{\mathcal O}(q^2)$\\
$\Q_{3,q}^{(6)}\to -2 F_A^{q/N} \Big(\op_7^N- \frac{m_N}{m_\chi} \op_9^N\Big)+{\mathcal O}(q^2)$\\
$\Q_{4,q}^{(6)}\to -4 F_A^{q/N} \op_4^N+F_{P'}^{q/N}\op_6^N+{\mathcal O}(q^2)$\\
$\Q_{1}^{(7)}\to F_G^{N}\op_1^N+{\mathcal O}(q^2)$\\
$\Q_{2}^{(7)}\to -\frac{m_N}{m_\chi}F_G^{N}\op_{11}^N +{\mathcal O}(q^3)$\\
$\Q_{3}^{(7)}\to F_{\tilde G}^{N}\op_{10}^N+{\mathcal O}(q^3)$\\
$\Q_{4}^{(7)}\to \frac{m_N}{m_\chi}F_{\tilde G}^{N}\op_{6}^N  +{\mathcal O}(q^4)$\\
$\Q_{5,q}^{(7)}\to F_S^{q/N}\op_1^N+{\mathcal O}(q)$\\
$\Q_{6,q}^{(7)}\to -\frac{m_N}{m_\chi}F_S^{q/N}\op_{11}^N +{\mathcal O}(q^2)$\\
$\Q_{7,q}^{(7)}\to F_{P}^{q/N}\op_{10}^N  +{\mathcal O}(q^3)$\\
$\Q_{8,q}^{(7)}\to \frac{m_N}{m_\chi}F_{P}^{q/N}\op_{6}^N   +{\mathcal O}(q^4)$\\
$\Q_{9,q}^{(7)}\to  8 F_{T,0}^{q/N}\op_4^N+ {\mathcal O}(q^2)$\\
$\Q_{10,q}^{(7)}\to  -2 \frac{m_N}{m_\chi} F_{T,0}^{q/N}\op_{10}^N
+ 2 \big(F_{T,0}^{q/N}-F_{T,1}^{q/N}\big)\op_{11}^N
-8 F_{T,0}^{q/N}\op_{12}^N+ {\mathcal O}(q^3)$\\
\hline
\end{tabular}
\caption{Correspondence between the relativistic operators listed in
  Eqs.(\ref{eq:dim5}--\ref{eq:dim7}) and the non-relativistic Galilean
  invariant operators listed in Table~\ref{table:operators}.}
\label{table:operator_correspondence}
\end{center}
\end{table}

For the details of the expression to calculate the expected rate in a
DD experiment we refer, for instance, to Section 2 of
\cite{sogang_scaling_law_nr}.  Once the coefficients of the
non--relativistic Hamiltonian in Eq.~(\ref{eq:H}) are obtained, for a
given recoil energy imparted to the target the differential rate for
the WIMP--nucleus scattering process is given by:

\be
\frac{d R_{\chi T}}{d E_R}(t)=\sum_T N_T\frac{\rho_{\mbox{\tiny WIMP}}}{m_{\mbox{\tiny WIMP}}}\int_{v_{min}}d^3 v_T f(\vec{v}_T,t) v_T \frac{d\sigma_T}{d E_R},
\label{eq:dr_de}
\ee

\noindent where $\rho_{\mbox{\tiny WIMP}}$ is the local WIMP mass density in the
neighborhood of the Sun, $N_T$ the number of the nuclear targets of
species $T$ in the detector (the sum over $T$ applies in the case of
more than one target), while

\be
\frac{d\sigma_T}{d E_R}=\frac{2 m_T}{4\pi v_T^2}\left [ \frac{1}{2 j_{\chi}+1} \frac{1}{2 j_{T}+1}|\mathcal{M}_T|^2 \right ],
\label{eq:dsigma_de}
\ee

\noindent with:

\begin{equation}
  \frac{1}{2 j_{\chi}+1} \frac{1}{2 j_{T}+1}|\mathcal{M}_T|^2=
  \frac{4\pi}{2 j_{T}+1} \sum_{\tau=0,1}\sum_{\tau^{\prime}=0,1}\sum_{k} R_k^{\tau\tau^{\prime}}\left [c^{\tau}_j,(v^{\perp}_T)^2,\frac{q^2}{m_N^2}\right ] W_{T k}^{\tau\tau^{\prime}}(y).
\label{eq:squared_amplitude}
\end{equation}

\noindent In the above expression the squared amplitude
$|\mathcal{M}_T|^2$ is summed over initial and final spins, the
$R_k^{\tau\tau^{\prime}}$'s are WIMP response functions which depend
on the couplings $c^{\tau}_j$ as well as the transferred momentum
$\vec{q}$, while:

\begin{equation}
(v^{\perp}_T)^2=v^2_T-v_{min}^2,
\label{eq:v_perp}
\end{equation}

\noindent and:

\begin{equation}
v_{min}^2=\frac{q^2}{4 \mu_{T}^2}=\frac{m_T E_R}{2 \mu_{T}^2},
\label{eq:vmin}
\end{equation}

\noindent represents the minimal incoming WIMP speed required to
impart the nuclear recoil energy $E_R$. Moreover, in equation
(\ref{eq:squared_amplitude}) the $W^{\tau\tau^{\prime}}_{T k}(y)$'s
are nuclear response functions and the index $k$ represents different
effective nuclear operators, which, under the assumption that the
nuclear ground state is an approximate eigenstate of $P$ and $CP$, can
be at most eight: following the notation in \cite{haxton1,haxton2},
$k$=$M$, $\Phi^{\prime\prime}$, $\Phi^{\prime\prime}M$,
$\tilde{\Phi}^{\prime}$, $\Sigma^{\prime\prime}$, $\Sigma^{\prime}$,
$\Delta$, $\Delta\Sigma^{\prime}$. The $W^{\tau\tau^{\prime}}_{T
  k}(y)$'s are function of $y\equiv (qb/2)^2$, where $b$ is the size
of the nucleus. For the target nuclei $T$ used in most DD
experiments the functions $W^{\tau\tau^{\prime}}_{T k}(y)$,
calculated using nuclear shell models, have been provided in
Refs.~\cite{haxton2,catena}. Details about the definitions of both the
functions $R_k^{\tau\tau^{\prime}}$'s and $W^{\tau\tau^{\prime}}_{T
  k}(y)$'s can be found in \cite{haxton2}.  In particular, $W_M$
corresponds to the standard Spin--Independent (SI) interaction, while
$W_{\Sigma^{\prime\prime}}+W_{\Sigma^{\prime}}$ (with
$W_{\Sigma^{\prime}}\simeq 2 W_{\Sigma^{\prime\prime}}$) to the
standard Spin--Dependent one (SD). Using the
decomposition:

\be
R_k^{\tau\tau^{\prime}}=R_{0k}^{\tau\tau^{\prime}}+R_{1k}^{\tau\tau^{\prime}} (v^{\perp}_T)^2=R_{0k}^{\tau\tau^{\prime}}+R_{1k}^{\tau\tau^{\prime}}\left (v_T^2-v_{min}^2\right ),
\label{eq:r_decomposition}
\ee

\noindent the correspondence between each term of the Non Relativistic
(NR) effective interaction in~(\ref{eq:H}) and the
$W^{\tau\tau^{\prime}}_{T k}(y)$ nuclear response functions is
summarized in Table \ref{table:eft_summary}.

\begin{table}[t]
\begin{center}
{\begin{tabular}{@{}|c|c|c|c|c|c|@{}}
\hline
NR coupling  &  $R^{\tau \tau^{\prime}}_{0k}$  & $R^{\tau \tau^{\prime}}_{1k}$ & coupling  &  $R^{\tau \tau^{\prime}}_{0k}$  & $R^{\tau \tau^{\prime}}_{1k}$ \\
\hline
$1$  &   $M(q^0)$ & - & $3$  &   $\Phi^{\prime\prime}(q^4)$  & $\Sigma^{\prime}(q^2)$\\
$4$  & $\Sigma^{\prime\prime}(q^0)$,$\Sigma^{\prime}(q^0)$   & - & $5$  &   $\Delta(q^4)$  & $M(q^2)$\\
$6$  & $\Sigma^{\prime\prime}(q^4)$ & - & $7$  &  -  & $\Sigma^{\prime}(q^0)$\\
$8$  & $\Delta(q^2)$ & $M(q^0)$ & $9$  &  $\Sigma^{\prime}(q^2)$  & - \\
$10$  & $\Sigma^{\prime\prime}(q^2)$ & - & $11$  &  $M(q^2)$  & - \\
$12$  & $\Phi^{\prime\prime}(q^2)$,$\tilde{\Phi}^{\prime}(q^2)$ & $\Sigma^{\prime\prime}(q^0)$,$\Sigma^{\prime}(q^0)$ & $13$  & $\tilde{\Phi}^{\prime}(q^4)$  & $\Sigma^{\prime\prime}(q^2)$ \\
$14$  & - & $\Sigma^{\prime}(q^2)$ & $15$  & $\Phi^{\prime\prime}(q^6)$  & $\Sigma^{\prime}(q^4)$ \\
\hline
\end{tabular}}
\caption{Nuclear response functions corresponding to each coupling,
  for the velocity--independent and the velocity--dependent components
  parts of the WIMP response function, decomposed as in
  Eq.(\ref{eq:r_decomposition}).  In parenthesis is the power of
  $q$=$|\vec{q}|$ in the WIMP response function.
  \label{table:eft_summary}}
\end{center}
\end{table}

Finally, $f(\vec{v}_T)$ is the
WIMP velocity distribution, for which we assume a standard isotropic
Maxwellian at rest in the Galactic rest frame truncated at the escape
velocity $u_{esc}$, and boosted to the Lab frame by the
velocity of the Earth. So for the former we assume:

\begin{eqnarray}
  f(\vec{v}_T,t)&=&\frac{1}{N}\left(\frac{3}{ 2\pi v_{rms}^2}\right )^{3/2}
  e^{-\frac{3|\vec{v}_T+\vec{v}_E|^2}{2 v_{rms}^2}}\Theta(u_{esc}-|\vec{v}_T+\vec{v}_E(t)|)\\
  N&=& \left [ \erf(z)-\frac{2}{\sqrt{\pi}}z e^{-z^2}\right ]^{-1},  
  \label{eq:maxwellian}
  \end{eqnarray}

\noindent with $z=3 u_{esc}^2/(2 v_{rms}^2)$. In the isothermal sphere
model hydrothermal equilibrium between the WIMP gas pressure and
gravity is assumed, leading to $v_{rms}$=$\sqrt{3/2}v_0$ with $v_0$
the galactic rotational velocity. The yearly modulation effect is due
to the time dependence of the Earth's speed with respect to the
Galactic frame:

\begin{equation}
|\vec{v}_E(t)|=v_{Sun}+v_{orb}\cos\gamma \cos\left [\frac{2\pi}{T_0}(t-t_0)
  \right ],
\label{eq:modulation}
  \end{equation}

\noindent where $\cos\gamma\simeq$0.49 accounts for the inclination of
the ecliptic plane with respect to the Galactic plane, $T_0$=1 year,
$t_0$=2 June, $v_{orb}$=2$\pi r_{\oplus}/(T_0)\simeq$ 29 km/sec
($r_{\oplus}$=1 AU, neglecting the small eccentricity of the Earth's
orbit around the Sun) while $v_{Sun}$=$v_0$+12, accounting for a
peculiar component of the solar system with respect to the galactic
rotation. For the two parameters $v_0$ and $u_{esc}$ we take $v_0$=220
km/sec \cite{v0_koposov} and $u_{esc}$=550 km/sec \cite{vesc_2014}.
In the isothermal model the time dependence of
Eq. (\ref{eq:modulation}) induces an expected rate with the functional
form $S(t)=S_0+S_m \cos(2\pi/T-t_0)$, with $S_m>0$ at large values of
$v_{min}$ and turning negative when $v_{min}\lsim$ 200 km/s.  

The expected rate in a given visible energy bin $E_1^{\prime}\le
E^{\prime}\le E_2^{\prime}$ of a DD experiment is
finally given by:

\begin{eqnarray}
R_{[E_1^{\prime},E_2^{\prime}]}(t)&=&MT_{exp}\int_{E_1^{\prime}}^{E_2^{\prime}}\frac{dR}{d
  E^{\prime}}(t)\, dE^{\prime} \label{eq:start}\\
 \frac{dR}{d E^{\prime}}(t)&=&\sum_T \int_0^{\infty} \frac{dR_{\chi T}(t)}{dE_{ee}}{\cal
   G}_T(E^{\prime},E_{ee})\epsilon(E^{\prime})\label{eq:start2}\,d E_{ee} \\
E_{ee}&=&Q(E_R) E_R \label{eq:start3},
\end{eqnarray}

\noindent with $\epsilon(E^{\prime})\le 1$ the experimental
efficiency/acceptance. In the equations above $E_R$ is the recoil
energy deposited in the scattering process (indicated in keVnr), while
$E_{ee}$ (indicated in keVee) is the fraction of $E_R$ that goes into
the experimentally detected process (ionization, scintillation, heat)
and $Q(E_R)$ is the quenching factor, ${\cal
  G_T}(E^{\prime},E_{ee}=Q(E_R)E_R)$ is the probability that the
visible energy $E^{\prime}$ is detected when a WIMP has scattered off
an isotope $T$ in the detector target with recoil energy $E_R$, $M$ is
the fiducial mass of the detector and $T_{exp}$ the live--time
exposure of the data taking.

In particular, in each visible energy bin DAMA is sensitive to the
yearly modulation amplitude $S_m$, defined as the cosine transform of
$R_{[E_1^{\prime},E_2^{\prime}]}(t)$:

\begin{equation}
S_{m,[E_1^{\prime},E_2^{\prime}]}\equiv \frac{2}{T_0}\int_0^{T_0}
\cos\left[\frac{2\pi}{T_0}(t-t_0)\right]R_{[E_1^{\prime},E_2^{\prime}]}(t)dt,
\label{eq:sm}
\end{equation}  

\noindent while other experiments put upper bounds on the time average
$S_0$:

\begin{equation}
S_{0,[E_1^{\prime},E_2^{\prime}]}\equiv \frac{1}{T_0}\int_0^{T_0}
R_{[E_1^{\prime},E_2^{\prime}]}(t)dt.
\label{eq:s0}
\end{equation}  

The expressions above can be recast in the form~\cite{generalized_halo_indep}:

\begin{equation}
  S_{m,0,[E_1^{\prime},E_2^{\prime}]}=\int_0^{\infty} {\cal R}(v)\eta_{m,0}(v)\, dv
  \label{eq:eta_factorization}
\end{equation}  

\noindent with ${\cal R}(v)$ a response function which contains the
dependence on the particle and nuclear physics while:

\begin{eqnarray}
  \eta_0(v)&=& \frac{1}{T_0}\int_0^{T_0}
\eta(v,t)dt,\\
  \eta_m(v)&=&\frac{2}{T_0}\int_0^{T_0}
  \cos\left[\frac{2\pi}{T_0}(t-t_0)\right] \eta(v,t)\,dt,\label{eq:eta_m}\\
    \eta(v,t)&=&\int_{v}^{\infty}\frac{f(v^{\prime},t)}{v^{\prime}}\,dv^{\prime},
\label{eq:eta_tilde}  
\end{eqnarray}

\noindent are halo functions that contain the dependence on
astrophysics.

In the present paper we will systematically consider the possibility
that the WIMP--nucleus interaction is driven by one of the effective
couplings $\C^{(d)}$ of Eqs.~(\ref{eq:dim5}--\ref{eq:dim7}).

\section{Analysis}
\label{sec:analysis}

The DAMA collaboration has released modulation amplitudes
$S_{m,k}^{exp}\equiv S_{m,[E_k^{\prime},E_{k+1}^{\prime}]}$, with
uncertainties $\sigma_k$, (corresponding to the predictions of
Eq.(\ref{eq:sm})) in the visible energy range 1 keVee$ < E^{\prime}<$
20 keVee in 0.5 keVee energy bins for a total exposure $\simeq$ 2.46
ton year, corresponding to the combination of DAMA/NaI
\cite{dama_1998}, DAMA/LIBRA--phase1 \cite{dama_2008,dama_2010} and
DAMA/LIBRA--phase2 \cite{dama_2018}.  In our analysis we will assume
constant quenching factors $Q$=0.3 for sodium, $Q$=0.09 for iodine and
a Gaussian energy resolution ${\cal
  G}(E^{\prime},E_{ee})=Gauss(E^{\prime}|E_{ee},\sigma)=1/(\sqrt{2\pi}\sigma)exp(-(E^{\prime}-E_{ee})/2\sigma^2)$
with $\sigma$ = 0.0091 (E$_{ee}$/keVee) + 0.448 $\sqrt{E_{ee}}$/keVee
in keVee. To compare the theoretical predictions to the experimental
data, for each coupling $\C^{(d)}$ we consider 14 energy bins, of 0.5
keVee width, from 1 keVee to 8 keVee, and one high--energy control bin
from 8 keVee to 16 keVee ($[E_k^{\prime},E_{k+1}^{\prime}]$,
$k=1,...,15$ ).  We perform our $\chi^2$ test constructing the
quantity:

\begin{equation}
\chi^2(m_{\chi},\tilde{\Lambda},r)=\sum_{k=1}^{15} \frac{\left
  [S_{m,k}-S^{exp}_{m,k}(m_{\chi},\tilde{\Lambda},r) \right ]^2}{\sigma_k^2},
  \label{eq:chi2}
  \end{equation}

\noindent and minimize it as a function of
$(m_{\chi},\tilde{\Lambda},r)$ (WIMP--photon and WIMP--gluon
interactions do not depend on the $r$ parameter).

\begin{figure}
\begin{center}
  \includegraphics[width=0.6\columnwidth]{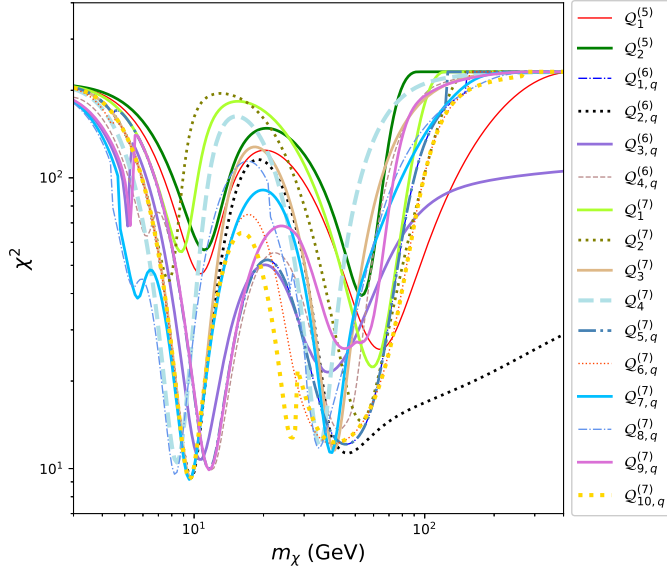}
\end{center}
\caption{Minimum of the $\chi^2$ of Eq.(\ref{eq:chi2}) at fixed WIMP
  mass $m_\chi$ as a function of $m_\chi$ for each of the different
  WIMP-nucleus interactions of
  Eqs.~(\ref{eq:dim5}--\ref{eq:dim7}).
\label{fig:chi2_m}}
\end{figure}
\begin{table}[ht!]
\begin{center}
\begin{tabular}{|M{1cm}|M{3cm}|M{2cm}|M{2.2cm}|M{2cm}|} \hline
\rule[-12.5pt]{0pt}{30pt}  $\mathbf{\cal{Q}}$  & $\mathbf{m_{\chi,min}}$ \textbf{(GeV)} & $\mathbf{r_{\chi,min}}$ & $\mathbf{\tilde\Lambda_{min}~(\mbox{\bf GeV})}$ & $\mathbf{\chi^2_{min}}$ \\\hline
  \multirow{ 2}{*}{${\cal Q}_1^{(5)}$}     &  63.95 & -- &  121.44 & 25.70 \\ 
                 &  10.62 & -- &  101.16 & 46.74 \\\hline
   \multirow{ 2}{*}{${\cal Q}_2^{(5)}$}      &  53.29 & -- &  50770.1 & 39.36 \\ 
                 &  11.09 & -- & 49155.7 & 56.45 \\\hline
 \multirow{ 2}{*}{${\cal Q}_{1,q}^{(6)}$}      &  11.62 & -0.88 &  311.62 &  9.94\\ 
                 &  45.55 & -0.87 & 516.67 & 12.11 \\\hline
   \multirow{ 2}{*}{${\cal Q}_{2,q}^{(6)}$}     &  9.71 & -1.14 &  29.55 &  9.28\\ 
                 &  46.24 & -0.58 & 82.53 &  11.28\\\hline  
  \multirow{ 2}{*}{${\cal Q}_{3,q}^{(6)}$}      &  10.65 & 3.30 & 2.20  &  10.70\\ 
                 &  38.13 & -5.43 & 8.06  & 21.45\\\hline 
  \multirow{ 2}{*}{${\cal Q}_{4,q}^{(6)}$}      &  11.93 & 3.17 & 62.68  &  9.90\\ 
                 &  42.43 & 2.93 & 82.26  &  13.58\\\hline
  \multirow{ 2}{*}{${\cal Q}_1^{(7)}$}      &  59.46 & -- & 60.50  &  22.32\\ 
                 &  8.77 & -- & 35.92  & 55.54 \\\hline
  \multirow{ 2}{*}{${\cal Q}_2^{(7)}$}   &  53.02 &  -- & 5.73   &  14.58\\ 
                 &  7.74 & -- &  3.60 & 44.08 \\\hline
  \multirow{ 2}{*}{${\cal Q}_3^{(7)}$}   &  9.63 &  -- & 3.80   &  9.26\\ 
                 &  40.42 & -- &  6.96 & 11.53 \\\hline
  \multirow{ 2}{*}{${\cal Q}_{4,q}^{(7)}$}   &  8.40 &  -- & 0.384   &  10.49\\ 
                 &  35.05 & -- &  0.712 & 11.93 \\\hline
  \multirow{ 2}{*}{${\cal Q}_{5,q}^{(7)}$}   &  11.62&  -0.20 & 5.26  &  9.94\\ 
                 &  45.55 & -0.20 &  7.38  & 12.11 \\\hline
  \multirow{ 2}{*}{${\cal Q}_{6,q}^{(7)}$}   &  9.57 & -0.20  & 1.12   & 9.16\\ 
                 &  40.12 &  -0.20 &  2.64   & 12.16  \\\hline
  \multirow{ 2}{*}{${\cal Q}_{7,q}^{(7)}$}   &  9.56 &  0.40 &  2.24  & 9.16\\ 
                 &  39.20 & 0.39  & 4.22  &  11.26\\\hline
  \multirow{ 2}{*}{${\cal Q}_{8,q}^{(7)}$}   &  8.27 &  0.36 &  0.366  & 9.52\\ 
                 &  34.71 & 0.14  &  1.44  &  11.70\\\hline
  \multirow{ 2}{*}{${\cal Q}_{9,q}^{(7)}$}   &  11.66 &  4.81 &  3.73  & 9.88\\ 
                 &  44.94 & 3.65  & 3.77  &  25.84\\\hline
  \multirow{ 2}{*}{${\cal Q}_{10,q}^{(7)}$}   &  9.57 &  -0.35 &  2.13  & 9.16\\ 
                 &  40.78 & -0.32  &  6.59  &  12.30\\\hline
\end{tabular}
\caption{Absolute and local minima of the $\chi^2$ defined in
  Eq.(\ref{eq:chi2}) for each of the relativistic models ${\cal
    Q}_{a,q}^{(d)}$ and ${\cal Q}_{b}^{(d)}$ of
  Eqs.~(\ref{eq:dim5}--\ref{eq:dim7}).}
\label{table:best_fit_values}
\end{center}
\end{table}

\begin{figure}
\begin{center}
  \includegraphics[width=0.6\columnwidth]{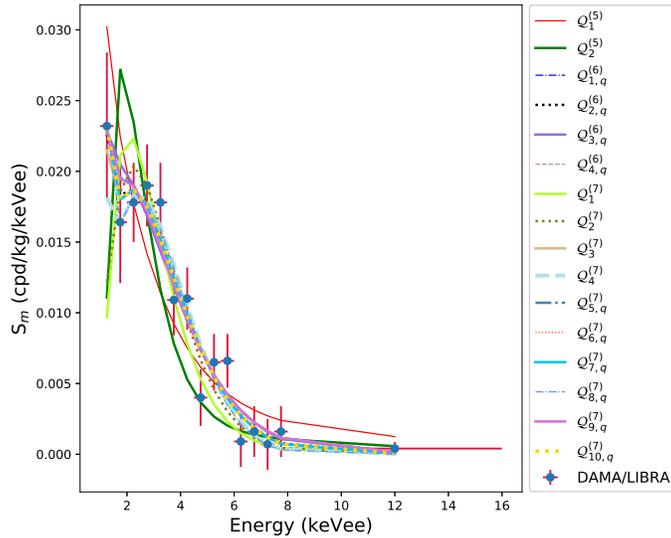}
\end{center}
\caption{DAMA modulation amplitudes as a function of the measured
  ionization energy $E_{ee}$ for the absolute minima of
  Table~\ref{table:best_fit_values} for each of the effective models. of
  Eqs.~(\ref{eq:dim5}--\ref{eq:dim7}). The points with
  error bars correspond to the combined data of DAMA/NaI
  \cite{dama_1998}, DAMA/LIBRA--phase1 \cite{dama_2008,dama_2010} and
  DAMA/LIBRA--phase2 \cite{dama_2018}.
\label{fig:e_sm}}
\end{figure}

\begin{figure}
\begin{center}
  \includegraphics[width=0.6\columnwidth]{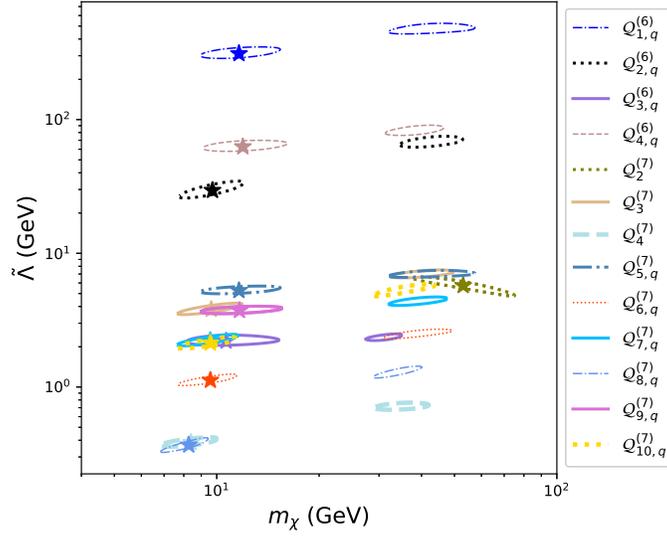}
\end{center}
\caption{ Contour plots of $\chi^2$-$\chi_{min}^2$=$n^2$ with $n=5$
  (5$\sigma$ regions) in the $m_{\chi}$--$\tilde{\Lambda}$ plane
  (minimized with respect to the $r$ parameter introduced in
  Eq.~(\ref{eq:r}), when applicable) for each of the 13 WIMP-nucleus
  interactions that yield an acceptable $\chi^2$. The stars
  correspond to the absolute minima of
  Table~\ref{table:best_fit_values}.
\label{fig:m_lambda}}
\end{figure}

In Fig. \ref{fig:chi2_m} we show the result of such minimization at
fixed WIMP mass $m_{\chi}$. From such figure one can see that for each
coupling $\C^{(d)}$ two local minima are obtained.  The details of
such minima are provided in Table~\ref{table:best_fit_values}, and the
ensuing predictions for the modulation amplitudes in the absolute
minima of each model are compared to those measured by DAMA in
Fig~\ref{fig:e_sm}. Moreover, the contour plots of
$\chi^2$-$\chi_{min}^2$=$n^2$ with $n=5$ (5$\sigma$ regions) in the
$m_{\chi}$--$\tilde{\Lambda}$ plane (minimized with respect to the $r$
parameter introduced in Eq.~(\ref{eq:r}), when applicable) are also
provided in Fig.~\ref{fig:m_lambda} for each of the 13 WIMP-nucleus
interactions that yield an acceptable $\chi^2$ in the absolute minimum
(in the following discussion we refer to a good fit when the
$p$--value is larger than 0.05, i.e.  $\chi^2\lsim$21 for 15-3 d.o.f. and
$\chi^2\lsim$22 for 15-2 d.o.f.). In the same figure the stars
correspond to the absolute minima of
Table~\ref{table:best_fit_values}.

The results summarized in Table \ref{table:best_fit_values} can be
understood in terms of a combination of kinematics, mainly driven by
the cosine transform of the halo function defined in
Eq.~(\ref{eq:eta_m}), and dynamics, determined instead by the
correspondence between each relativistic model coupling and its
non-relativistic limit, as summarized in
Table~\ref{table:operator_correspondence}.

\begin{figure}
\begin{center}
  \includegraphics[width=0.45\columnwidth]{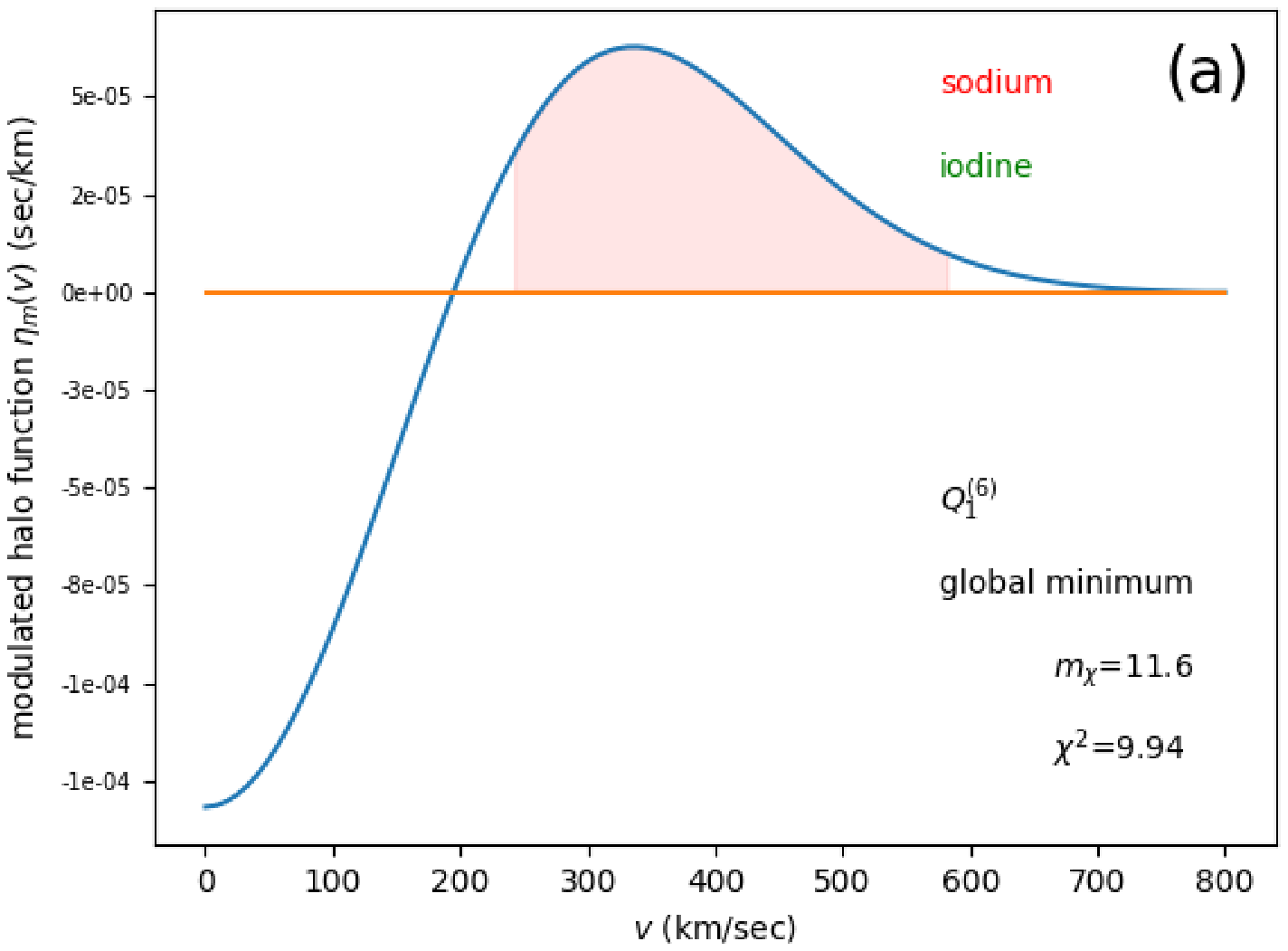}
  \includegraphics[width=0.45\columnwidth]{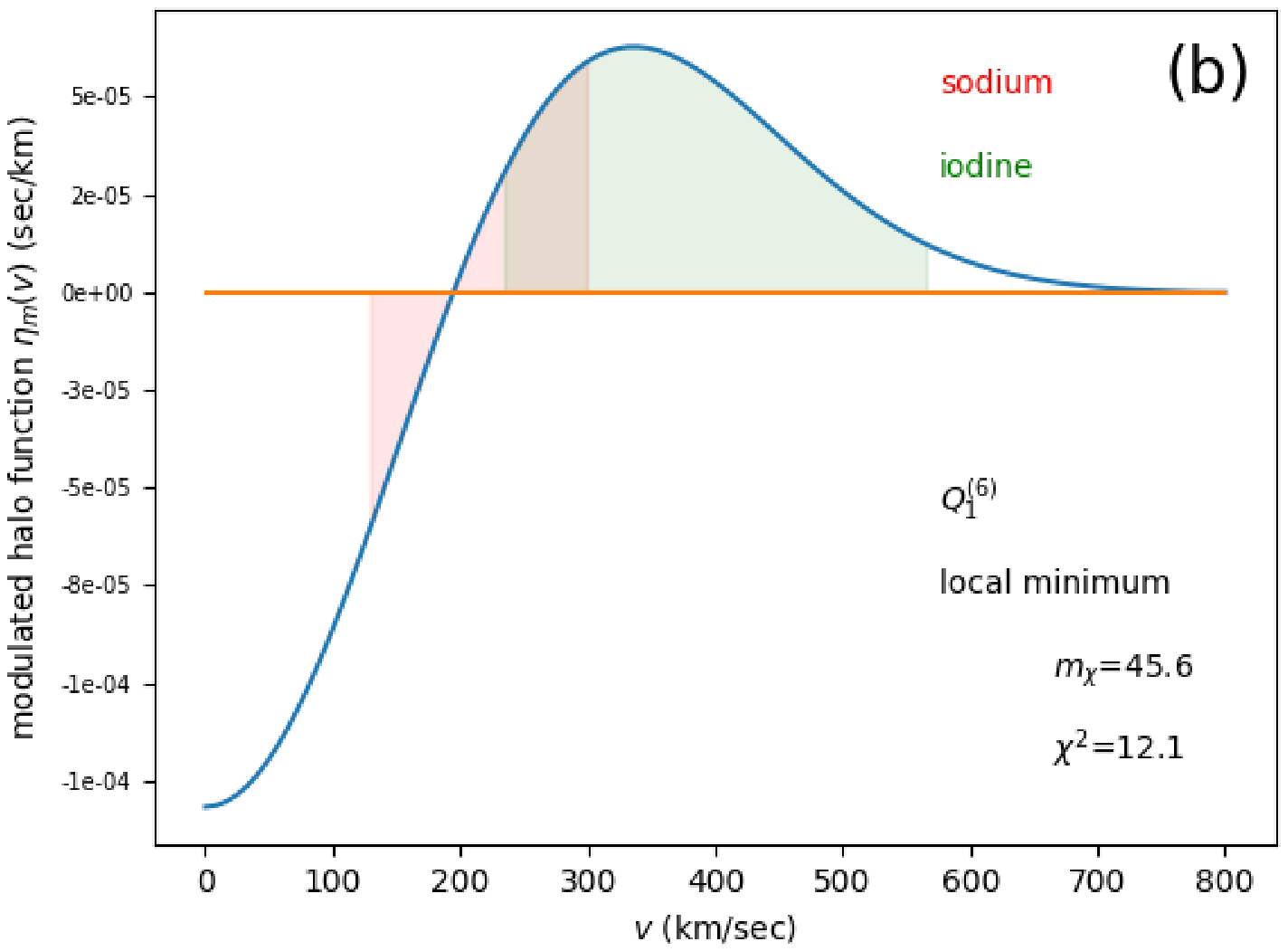}  
\end{center}
\caption{Cosine transform $\eta_m$ of the halo function $\eta(v,t)$
  defined in Eq.~\ref{eq:eta_m}, and entering the expected amplitudes
  parameterization of Eq.~\ref{eq:eta_factorization}. In the same
  figures the shaded areas represent the range of $v$ contributing to
  the DAMA modulated amplitudes between 1 keVee and 6 keVee for
  WIMP-sodium scattering (red) and WIMP--iodine scattering (green) for
  the interaction term $\Q^{(6)}_{1,q}$ and and in the minima of
  Table~\ref{table:best_fit_values}. {\bf (a)}: low--mass minimum;
  {\bf (b)}: high--mass minimum.
\label{fig:halo_function}}
\end{figure}

To illustrate the role of the halo function in the kinematics of the
scattering process, in Fig.~\ref{fig:halo_function} we plot the
function $\eta_m(v)$. In particular, if in
Eq.~(\ref{eq:eta_factorization}) the response function ${\cal R}$ does
not show a strong energy dependence induced by the nuclear form
factor, or, as in the case of the $\op_7$ or $\op_{14}$ NR operators,
an explicit dependence on the velocity $v$ through a non--vanishing
$R_{1k}^{\tau\tau^{\prime}}$ (see Eq.~(\ref{eq:r_decomposition})) it
is the $\eta_m(v)$ halo function that drives the energetic spectral
shape of the modulation amplitude, and the latter can well reproduce
the maximum observed above 2 keVee (see Fig.~\ref{fig:e_sm}), once the
$v$ range close to $\simeq$340 km/sec (for our choice of astrophysical
parameters) is mapped to recoil energies around $\simeq$ 3 keVee
through Eq.~(\ref{eq:vmin}). Indeed, with a few exceptions discussed
below, this is the mechanism that allows to achieve most of the
acceptable $\chi^2$'s listed in
Table~\ref{table:best_fit_values}. However, due to the different
kinematics of the two processes, the energy range of the observed
maximum corresponds to two different values of $m_{\chi}$ for
WIMP--sodium and WIMP--iodine scattering. This explains the systematic
presence of two minima in the $\chi^2$.

Moreover, in Fig.~\ref{fig:halo_function} the shaded areas
represent the range of $v$ contributing to the DAMA modulated
amplitudes between 1 keVee and 6 keVee for WIMP-sodium scattering
(red) and WIMP--iodine scattering (green) in the specific example of
the interaction term $\Q^{(6)}_{1,q}$ and in the the corresponding two
minima of Table~\ref{table:best_fit_values} (such $v$ ranges are
defined as the intervals where in Eq.~(\ref{eq:eta_factorization}) the
response function ${\cal R}(v)$ is different from zero).  In
particular, in Fig.~\ref{fig:halo_function}(a), where the shaded areas
are shown for the low--mass minimum, one can see that scattering off
iodine is driven to $v$ values very close to the escape velocity ( for
this reason it is not visible). Indeed, in such case scattering off
iodine is kinematically accessible only below 2 keVee, while above
that energy sodium alone contributes to the expected signal. On the
other hand, Fig.~\ref{fig:halo_function}(b) shows that in the
high--mass minimum both targets contribute. However, in this case
sodium is pushed down to low $v$ values, where the the
modulation amplitude is suppressed.

As a consequence of the above discussion, the goodness--of--fit of the
DAMA modulation amplitudes at low WIMP mass is sensitive to both
WIMP--iodine and WIMP--sodium scattering, and so to the scaling
between the two corresponding cross sections~\cite{freese_2018,
  dama_2018_sogang, cosinus}. Indeed, in such case, provided that such
hierarchy is not too large, a good fit can be attained since the
energy dependence of the modulation amplitude is mostly driven by the
$\eta_m(v)$ halo function and is only slightly modified above 2 keVee
by the sodium form factor, which has a mild additional energy
dependence due to the small size target and the sensitivity to smaller
recoil energies at fixed electron--equivalent ones compared to iodine
due to the larger quenching factor.  On top of that, below 2 keVee
WIMP--iodine scattering can provide a nice fit to the modulation
amplitude measured by DAMA in the first energy bin: indeed, it is for
this reason that in Table~\ref{table:best_fit_values} most of the
absolute minima correspond to the low WIMP mass solution.  Among the
different NR operators of Table~\ref{table:operators} the case of a
standard SI cross section, which corresponds to the NR operator
$\CO_1$, shows the largest hierarchy between the WIMP--iodine and
WIMP--sodium cross sections, since they depend on the nuclear response
function $W^{\tau\tau^{\prime}}_M$ (see Table~\ref{table:eft_summary})
which scales with the mass numbers of the two targets. This makes it
hard to fit WIMP--sodium scattering above 2 keVee without
overestimating the WIMP--iodine process at lower energy unless the
non--relativistic couplings $c^n_j$/$c^p_j$ off neutrons and protons
are tuned to suppress the WIMP--iodine cross section, as pointed out
in~\cite{freese_2018, dama_2018_sogang, cosinus2}. In particular, by
combining the information contained in
Table~\ref{table:operator_correspondence} and
Table~\ref{table:eft_summary} one can notice that nine out of the
sixteen relativistic effective models of
Table~\ref{table:best_fit_values} take contribution from the
$W^{\tau\tau^{\prime}}_{TM}$ nuclear response function:
$\Q_{1}^{(5)}$, $\Q_{2}^{(5)}$, $\Q_{1}^{(6)}$, $\Q_{1}^{(6)}$,
$\Q_{1}^{(7)}$, $\Q_{2}^{(7)}$, $\Q_{5,q}^{(7)}$, $\Q_{6,q}^{(7)}$ and
$\Q_{10,q}^{(7)}$. In particular, four of such operators correspond to
effective interactions between the WIMP and gauge bosons (the photon
and the gluon) for which the ratio between the non--relativistic
couplings $c^n_j$/$c^p_j$ is fixed, and cannot be tuned to reduce the
$\chi^2$: $\Q_{1}^{(5)}$, $\Q_{2}^{(5)}$, $\Q_{1}^{(7)}$ and
$\Q_{2}^{(7)}$. Indeed, at low WIMP mass model $\Q_{1}^{(7)}$ has
$\chi^2$=22.32 and $\Q_{2}^{(7)}$ has $\chi^2$=44.08, while the
$\chi^2$ for models $\Q_{1}^{(5)}$ and $\Q_{2}^{(5)}$ is even higher,
since they have the additional feature that the $1/q^2$ propagator
further enhances the expected modulation amplitudes in the first two
energy bins (for such long--range interactions this implies that also
the $\chi^2$ at large WIMP masses is not acceptable).  On the other
hand the interactions $\Q_{1,q}^{(6)}$, $\Q_{2,q}^{(6)}$,
$\Q_{5,q}^{(7)}$, $\Q_{6,q}^{(7)}$ and $\Q_{10,q}^{(7)}$ allow to tune
the $r$ parameter to suppress the iodine contribution, leading to an
acceptable $\chi^2$. Finally, the remaining seven interactions
$\Q_{3}^{(6)}$, $\Q_{4}^{(6)}$, $\Q_{3}^{(7)}$, $\Q_{4}^{(7)}$,
$\Q_{7}^{(7)}$, $\Q_{8}^{(7)}$ and $\Q_{9}^{(7)}$ achieve a good
$\chi^2$ at small $m_{\chi}$ without any particular tuning because
they are all driven by SD type nuclear form factors (i.e.
$W^{\tau\tau^{\prime}}_{T\Sigma^{\prime\prime}}$ and/or
$W^{\tau\tau^{\prime}}_{T\Sigma^{\prime}}$) for which no large
hierarchy is observed in the first place between the WIMP--iodine and
the WIMP--sodium cross sections.

As far as the large--mass minima of Table~\ref{table:best_fit_values}
are concerned, as explained above they correspond to the situation
when WIMP--iodine scattering maps the maximum of the halo function
$\eta_m(v)$ at $v\simeq$ 340 km/sec to recoil energies $\simeq$ 3 keV.
In this case to get an acceptable fit the WIMP--sodium scattering
process is not relevant, since, as shown in
Fig.~\ref{fig:halo_function}(b), for such WIMP masses it is driven to
values of $v$ where its contribution to the modulation is
suppressed. As a consequence, in the high--mass minima the modulation
amplitudes are sensitive to WIMP--iodine scattering alone, while the
hierarchy between the WIMP--sodium and WIMP--iodine cross sections
does not play a significant role. In this case, however, another
effect kicks in: due to the large size and the small quenching of the
iodine target all the corresponding nuclear form factors
$W^{\tau\tau^{\prime}}_{Tk}$ show a much steeper energy dependence
compared to that of sodium.  This additional energy dependence in the
WIMP response function ${\cal R}$ alters the spectrum provided by the
$\eta_m(v)$ halo function alone spoiling the goodness-of-fit at large
WIMP masses, unless some mechanism is active to reduce it.  In the
models where an acceptable $\chi^2$ is achieved at large WIMP mass
this mechanism is provided by an explicit momentum suppression in the
WIMP response function $R_k^{\tau\tau^{\prime}}$ and/or (in the case
of WIMP--quark interactions driven by $W^{\tau\tau^{\prime}}_{TM}$) by
a tuning of the $c^n/c^p$ ratio.

In particular this explains why, out of the seven models with a
momentum--suppressed SD type interaction ($\Q_{3,q}^{(6)}$ and
$\Q_{9,q}^{(7)}$, $\Q_{4,q}^{(6)}$, $\Q_{3}^{(7)}$, $\Q_{4}^{(7)}$,
$\Q_{7,q}^{(7)}$ and $\Q_{8,q}^{(7)}$) all achieve a good $\chi^2$ at
large WIMP masses with the exception of $\Q_{3,q}^{(6)}$ and
$\Q_{9,q}^{(7)}$. It is worth noting that this implies that for
$\Q_{4,q}^{(6)}$ a tuning of the $r$ parameter to enhance the $\op_6$
contribution ($q^4$ suppressed) compared to $\op_4$ (with no momentum
suppression) is required to achieve a good $\chi^2$.  Indeed, the
$\op_4$ operator (i.e. a standard, SD interaction) has already been
shown to yield a bad fit at large $m_\chi$ to the DAMA
data~\cite{dama_2018_sogang}. This also explains why $\Q_{9,q}^{(7)}$,
that corresponds to the NR operator $\op_4$ alone (again, see
Table~\ref{table:operator_correspondence}) yields a bad $\chi^2$ . On
the other hand, the operator $\Q^{(6)}_{3,q}$ corresponds to a
combination of $\op_7$ and $\op_9$, with the latter suppressed at
large $m_{\chi}$ (see Table~\ref{table:operator_correspondence}) so
that in this case the contribution of $\op_7$ turn out to be
sizeable. However (see Table~\ref{table:eft_summary} and
Eq.~(\ref{eq:r_decomposition})) the WIMP response function
$R^{\tau\tau^{\prime}}_k$ of $\op_7$ shows in this case an explicit
dependence on $v^2$ that spoils the fit (the same was observed in the
fit of the $\op_7$ operator of Ref.~\cite{dama_2018_sogang}). For the
five models that depend on $W^{\tau\tau^{\prime}}_{TM}$ and that
correspond to effective interactions of the WIMP to quarks
($\Q_{1,q}^{(6)}$, $\Q_{2,q}^{(6)}$, $\Q_{5,q}^{(7)}$,
$\Q_{6,q}^{(7)}$ and $\Q_{10,q}^{(7)}$) besides momentum suppression,
when present, the ratio of the non--relativistic couplings
$c^n_j$/$c^p_j$ can be tuned to reduce the steepness of the iodine SI
form factor and achieve an acceptable $\chi^2$.  Such dependence on
$c^n/c^p$ is not present for the two WIMP--gluon effective
interactions $\Q_{1,q}^{(7)}$ and $\Q_{2,q}^{(7)}$: the former has no
momentum suppression either, and so yields a a bad $\chi^2$, while the
latter is momentum suppressed, so corresponds to an acceptable
goodness--of--fit.

The axial current in $\Q^{(6)}_{4,q}$, the pseudoscalar current in
$\Q^{(7)}_{7,q}$ and $\Q^{(7)}_{8,q}$ and the CP--odd gluonic current
in $\Q^{(7)}_{3,q}$ and $\Q^{(7)}_{4,q}$ deserve a few additional
comments since they receive contributions from light pseudoscalar
meson exchanges~\cite{bishara_2017} that can potentially modify the
modulation energy spectrum. In particular, this implies that the
Wilson coefficients of the NR effective theory can potentially acquire
an additional momentum dependence, since in
Table~\ref{table:operator_correspondence} one has~\cite{bishara_2017}:

\begin{eqnarray}
\label{eq:F_PP'}
F_{P,P'}^{q/N}(q^2)&=&\frac{m_N^2}{m_\pi^2+q^2} a_{\pi}^{q/N}+\frac{m_N^2}{m_\eta^2+q^2} a_{\eta}^{q/N}+b^{q/N}\label{eq:P_poles}
,
\\
\label{eq:F_tildeG}
F_{\tilde G}^{N}(q^2)&=&
\frac{-q^2}{m_\pi^2+q^2} a_{\tilde G,\pi}^{N}+\frac{-q^2}{m_\eta^2+q^2} a_{\tilde G,\eta}^{N}+b_{\tilde G}^{N},
\end{eqnarray}

\noindent with $m_\pi\simeq$ 135 MeV and $m_\eta\simeq$ 547 MeV the
pion and eta meson masses, and $a_{\pi}^{q/N}$, $a_{\eta}^{q/N}$,
$b^{q/N}$, $a_{\tilde G,\pi}^{N}$, $a_{\tilde G,\eta}^{N}$ and
$b_{\tilde G}^{N}$ constants that we calculate using the code
DirectDM~\cite{directdm} and that depend on the $r$ parameter.  In
particular, in light of the previous discussion such meson
propagators, as in the case of the long--range interactions
$\Q^{(5)}_1$ and $\Q^{(5)}_2$, could potentially spoil the fit to the
DAMA modulation amplitudes. However, what we observe is that the
effect of such poles is always mild. In fact one should first notice
that meson poles can modify the energy spectrum only for a pion
propagator, and only when the scattering rate is driven by
WIMP--iodine scattering, since in this case one has $q^2$ = $2 M_T
E_R\lsim$ 0.8 $m_\pi^2$ in the DAMA signal region $E_{ee}\lsim$ 6
keVee (the same quantity for sodium is $q^2 \lsim$ 0.04
$m_\pi^2$). Moreover, in the case of the CP--odd gluonic current of
$\Q^{(7)}_{3,q}$ and $\Q^{(7)}_{4,q}$ the constant term $b_{\tilde
  G}^{N}$ is LO and dominates the Wilson coefficient, with the poles
representing a small correction, as we observe numerically. Also, in
the case of $\Q^{(6)}_{4,q}$ the contribution of the term $F_{P'}
\op_6$, albeit, as pointed out previously, instrumental to improve the
fit at high WIMP mass, does not exceed $\simeq$ 10\% of the total
rate. Only in the case of the pseudoscalar current in $\Q^{(7)}_{7,q}$
and $\Q^{(7)}_{8,q}$ the constant term $b^{q/N}$ is NLO, so in this
case the terms proportional to the meson poles actually drive the
predicted rates. However both in $\Q^{(7)}_{7,q}$ and $\Q^{(7)}_{8,q}$
after hadronization~\cite{bishara_2017} the pion pole acquires an
isovector coupling ($c^p_\pi\simeq -c^n_\pi$) and the eta pole an
isoscalar one ($c^p_\eta\simeq c^n_\eta$) so that $F_{P,P'}^{q/N}
\rightarrow c^{\tau}$ =
$c^{1}_{\pi}/(m_\pi^2+q^2)+c^{0}_{\eta}/(m_\eta^2+q^2)$. Taking into
account Tables~\ref{table:operator_correspondence} and
\ref{table:eft_summary} one can see that in this case the scattering
amplitude is proportional to $q^n c^{\tau}c^{\tau^{\prime}}
W^{\tau\tau^{\prime}}_{T\Sigma^{\prime\prime}}$ = $q^n[ (c^{1}_\pi)^2
  W^{11}_{T\Sigma^{\prime\prime}}/(m_\pi^2+q^2)^2 + 2 c^{0}_\pi
  c^{1}_\pi
  W^{01}_{T\Sigma^{\prime\prime}}/((m_\pi^2+q^2)(m_\eta^2+q^2))+(c^{0}_\eta)^2
  W^{00}_{T\Sigma^{\prime\prime}})/(m_\eta^2+q^2)^2]$, with $n$=2 for
$\Q^{(7)}_{7,q}$ and $n$=4 for $\Q^{(7)}_{7,q}$. Such expression turns
out to have a mild momentum dependence because the three functions
$W^{11}_{T\Sigma^{\prime\prime}}(q)$,
$W^{01}_{T\Sigma^{\prime\prime}}(q)$ and
$W^{00}_{T\Sigma^{\prime\prime}}(q)$ show an increasing steepness in
their $q$ dependence that compensates the decreasing steepness of the
corresponding pole combinations $1/(m_\pi^2+q^2)^2$,
$1/((m_\pi^2-q^2)(m_\eta^2+q^2))$ and $1/(m_\eta^2+q^2)^2$. The bottom
line is that also in this case the overall momentum dependence does
not depart significantly from the other momentum--suppressed SD
interactions already discussed before, and so a good fit to the DAMA
data can be achieved.

Besides the goodness of fit, one should notice at this stage that the
values of $\tilde{\Lambda}_{min}$ in Table~\ref{table:best_fit_values}
should be checked for their consistency with the validity of the
effective theory.  A criterion for the validity of the EFT is to
interpret both scales $\tilde{\Lambda}_{up}$ = $\tilde{\Lambda}$ and
$\tilde{\Lambda}_{down} \equiv ({\C^{(d)}_{down}})^{-\frac{1}{d-4}}$ =
$\tilde{\Lambda}_{up} \times r^{-\frac{1}{d-4}}$ in terms of the
propagator $g^2/M_{*}^2$ with $g<\sqrt{4\pi}$ and $M_{*}>\mu_{scale}$,
where, in our analysis, we have fixed the boundary conditions of the
EFT at the scale $\mu_{scale}$ = 2 GeV. This is straightforward for
dimension--6 operators, while in the case of operators whose effective
coupling has dimension different from -2 only matching the EFT with
the full theory would allow to draw robust conclusions. In particular,
in this case $\tilde{\Lambda}_{up}$ and $\tilde{\Lambda}_{down}$ can
be interpreted in terms of the same propagator times the appropriate
power of a typical scale of the problem $\mu_{scale}^{\prime}$, which
depends on the ultraviolet completion of the EFT. To fix an order of
magnitude we choose to fix $\mu_{scale}^{\prime}$ = $\mu_{scale}$, so
that the bound
$min(\tilde{\Lambda}_{up},\tilde{\Lambda}_{down})>\mu_{scale}/(4\pi)^{1/(d-4)}$
can be derived. Using the values of $\tilde{\Lambda}_{min}$ and
$r_{min}$ in Table~\ref{table:best_fit_values} we obtain that such
bound is violated for the global minima of $\Q^{(7)}_4$ and
$\Q^{(7)}_{8,q}$ and the local minimum of $\Q^{(7)}_4$. However we
stress again that for operators with dimension different from 6 this
can only be assessed when a specific ultraviolet completion of the
effective theory is assumed.

\begin{figure}
\begin{center}
  \includegraphics[width=0.45\columnwidth]{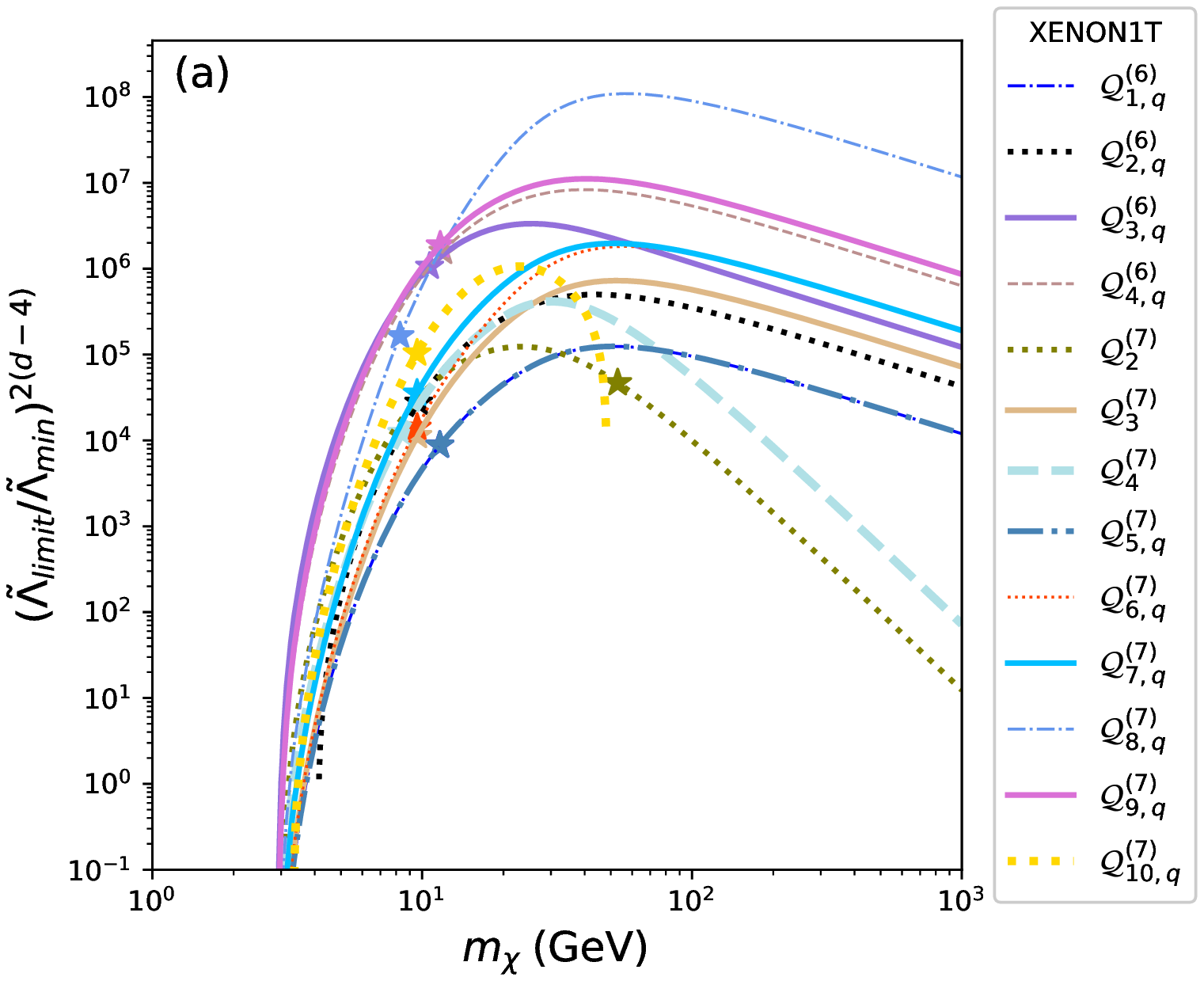}
  \includegraphics[width=0.45\columnwidth]{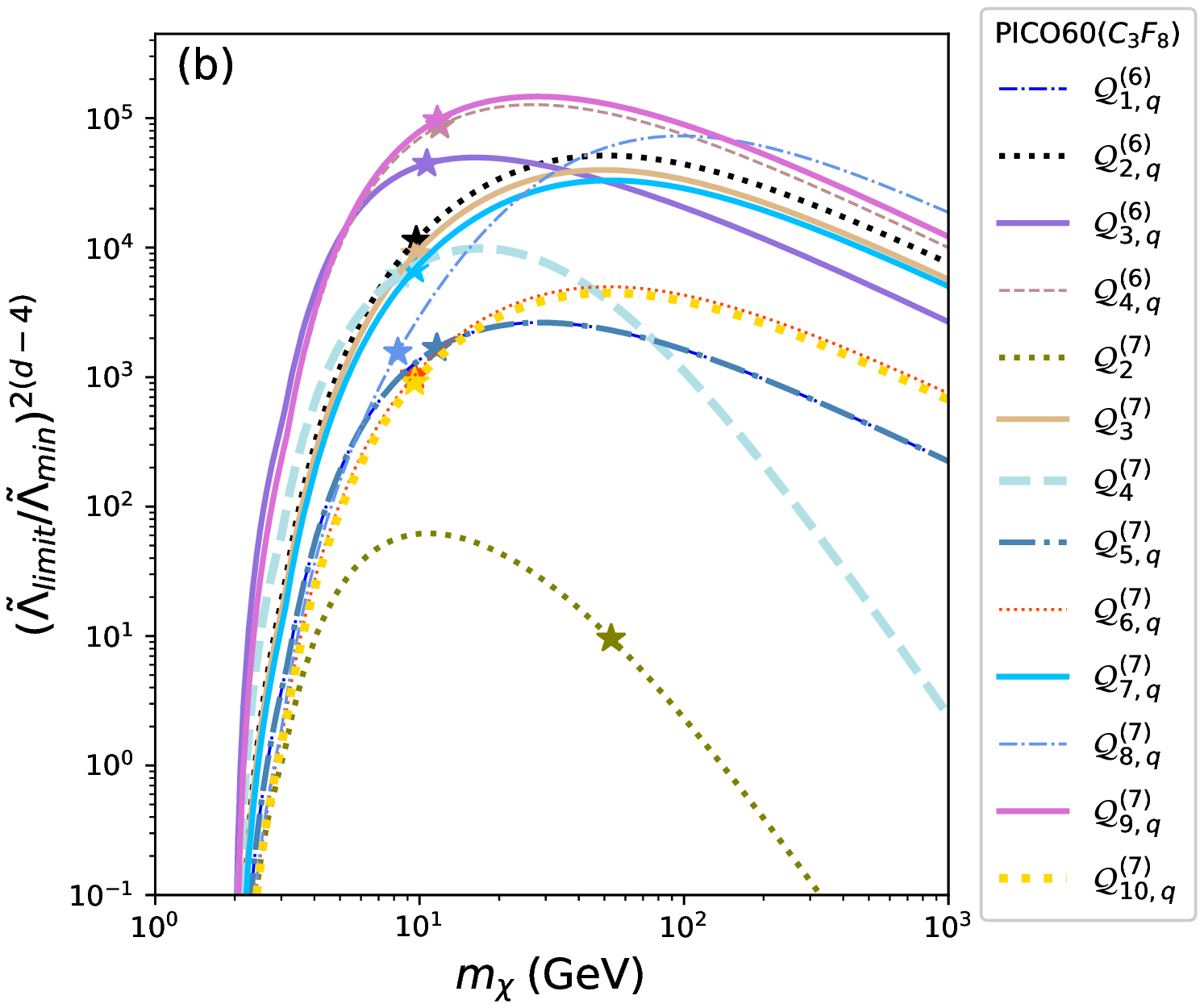}
\end{center}
\caption{Ratio
  $(\tilde{\Lambda}_{limit}/\tilde{\Lambda}_{min})^{2(d-4)}$ (with $d$
  the dimensionality of the operator) between the experimental lower
  bound $\tilde{\Lambda}_{limit}$ on the effective scale
  $\tilde{\Lambda}$ and its best--fit value $\tilde{\Lambda}_{min}$
  from Table~\ref{table:best_fit_values} as a function of the WIMP
  mass $m_{\chi}$ for each of the 13 relativistic effective models
  that yield an acceptable $\chi^2$.  Such combination corresponds for
  each model to the ratio between the expected number of events and
  the corresponding upper bound. {\bf (a)} $\tilde{\Lambda}_{min}$
  from XENON1T~\cite{xenon_2018}; {\bf (b)} $\tilde{\Lambda}_{min}$
  from PICO--60~\cite{pico60_2019}. For all models the ratio
  $r=\C^{(d)}_{down}/\C^{(d)}_{up}$ is fixed to its best--fit value
  from Table~\ref{table:best_fit_values}, and the stars indicate the
  corresponding best--fit value of the WIMP mass $m_{\chi}$.
\label{fig:bounds}}
\end{figure}

We conclude this section comparing in Fig.~\ref{fig:bounds} the
best--fit values $\tilde{\Lambda}_{min}$ from
Table~\ref{table:best_fit_values} for the effective scale
$\tilde{\Lambda}$ and the corresponding experimental lower bound
$\tilde{\Lambda}_{limit}$ from XENON1T~\cite{xenon_2018}
(Fig.~\ref{fig:bounds}(a)) and PICO--60~\cite{pico60_2019}
(Fig.~\ref{fig:bounds}(b)). In particular, for XENON1T we have assumed
7 WIMP candidate events in the range of 3PE $ \le S_1 \le $ 70PE, as
shown in Fig.~3 of Ref.~\cite{xenon_2018} for the primary
scintillation signal S1 (directly in Photo Electrons, PE), with an
exposure of 278.8 days and a fiducial volume of 1.3 ton of xenon.  We
have used the efficiency taken from Fig.~1 of~\cite{xenon_2018} and
employed a light collection efficiency $g_1$=0.055; for the light
yield $L_y$ we have extracted the best estimation curve with an
electric field of $90~{\rm V/cm}$ from Fig. 7 for
Ref.~\cite{xenon_2018_quenching}.  Moreover, we have modeled the
energy resolution combining a Poisson fluctuation of the observed
primary signal $S_1$ compared to $<S_1>$ = $g_1 L_y E_R$ and a
Gaussian response of the photomultiplier with
$\sigma_{PMT}=0.5$~\cite{xenon100_reponse}. On the other hand, for
PICO--60 we have combined the two runs obtained using a
$C_3F_8$~\cite{pico60_2019} target discussed in~\cite{pico60_2019}: a
1404 kg day exposure and threshold $E_{th}$=2.45, with 3 observed
candidate events and 1 event from the expected background, implying an
upper bound of 6.42 events at 90\%C.L.; a 1167 kg day exposure and
threshold $E_{th}$=3.3 keV, with zero observed candidate events and
negligible expected background, implying a 90\% C.L. upper bound of
2.3 events. For both runs we have assumed the nucleation probabilities
in Fig. 3 of \cite{pico60_2019}.

The result of the comparison between the best-fit minima of
Table~\ref{table:best_fit_values} and the XENON1T and PICO--60
constraints is shown in Fig.~\ref{fig:bounds} for each of the 13
models that yield an acceptable $\chi^2$. In particular, in that
figure we plot the combination
$(\tilde{\Lambda}_{limit}/\tilde{\Lambda}_{min})^{2(d-4)}$ (with $d$
the dimensionality of the operator) which corresponds to the ratio
between the expected number of events and the corresponding upper
bound. In both plots the ratio $r=\C^{(d)}_{down}/\C^{(d)}_{up}$ is
fixed to its best--fit value from Table~\ref{table:best_fit_values}
(when applicable), and the stars indicate the corresponding best--fit
value of the WIMP mass $m_{\chi}$.  From Fig.~(\ref{fig:bounds}) one
can see that for all the absolute minima of
Table~\ref{table:best_fit_values} the corresponding predicted number
of events exceeds by more than three orders of magnitude the upper
bound from XENON1T and/or PICO--60.

\section{Conclusions}
\label{sec:conclusions}

Model-independent approaches have become increasingly popular to
interpret Dark Matter search experiments, due to the lack of evidence
of any physics beyond the standard model from the LHC. However, in
spite of its potential relevance, a completely model--independent
assessment of the DAMA modulation result is still not available after
almost twenty years of its appearance. In particular, no DAMA analysis
in terms of a WIMP effective interaction with photons, gluons and
quarks was available so far, and, while the existing ones in terms of
the non--relativistic operators allowed by Galilean invariance do not
include Wilson coefficients with an explicit momentum dependence,
relativistic effective models can induce them. In order to fill this
gap in the present paper we have analyzed the DAMA/Libra--phase2
modulation result using a basis of 16 effective operators describing
the WIMP interaction with photons, gluons and quarks up to mass
dimension seven.  For each operator we have fixed the effective theory
at the scale of 2 GeV and parameterized the WIMP--quark interaction in
terms of two independent couplings $\C^{(d)}_{down}$ and
$\C^{(d)}_{up}$ common to up--type and down--type quarks.  We have
then discussed the fit to the DAMA--phase2 data in terms of the WIMP
mass $m_{\chi}$, the effective scale $\tilde{\Lambda}$ defined through
the parameterization $\C^{(d)}_{up}$=1/$\tilde{\Lambda}^{d-4}$ (with
$d$ the effective operator dimensionality) and the ratio $r$ =
$\C^{(d)}_{down}/\C^{(d)}_{up}$. For all the effective operators that
we considered we obtained two minima, one at low WIMP mass $\simeq$ 10
GeV and the other for $m_{\chi}\gsim$ 35 GeV. In particular, only nine
out of the ensuing thirty--two minima and three out of the sixteen
models do not have an acceptable goodness--of--fit. Two additional
models ($\Q^{(7)}_4$ and $\Q^{(7)}_{8,q}$) require values of the
effective scale below 2 GeV, i.e. the value that we use to fix the
Wilson parameters.

In both the low and high $m_{\chi}$ minima for $\Q^{(5)}_1$ and
$\Q^{(5)}_2$ (corresponding to a magnetic dipole and electric dipole
WIMP coupling) the photon propagator induces a steep rise of the
modulation amplitudes at low WIMP recoil energy incompatible to the
measured ones for both low and high WIMP masses.  In the remaining two
models with a bad $\chi^2$ at low $m_{\chi}$ the hierarchy between the
WIMP--iodine and the WIMP-sodium cross section is too large because it
is of the SI type (i.e. it scales with the atomic
squared mass) and cannot be reduced by tuning the $r$ parameter
($\Q^{(7)}_1$ and $\Q^{(7)}_2$). On the other hand, at high $m_{\chi}$
the iodine nuclear form factor has a too steep energy dependence for
$\Q^{(7)}_1$ and $\Q^{(7)}_{9,q}$, while model $\Q^{(6)}_{3,q}$ does
not yield an acceptable fit because the corresponding WIMP response
function depends explicitly on the velocity $v$. In all the other
twenty--three minima a good $\chi^2$ can be achieved. In particular in
the twelve low--WIMP mass minima the hierarchy between the
WIMP--iodine and the WIMP-sodium cross section is either naturally of
order one because of a SD type interaction
($\Q_{3}^{(6)}$, $\Q_{4}^{(6)}$, $\Q_{3}^{(7)}$, $\Q_{4}^{(7)}$,
$\Q_{7}^{(7)}$, $\Q_{8}^{(7)}$ and $\Q_{9}^{(7)}$) or can kept under
control by tuning the $r$ parameter ($\Q_{1,q}^{(6)}$,
$\Q_{2,q}^{(6)}$, $\Q_{5,q}^{(7)}$, $\Q_{6,q}^{(7)}$ and
$\Q_{10,q}^{(7)}$). Finally, in the eleven large--$m_{\chi}$ minima an
acceptable goodness--of--fit is achieved because the steepness of the
iodine nuclear form factor is either compensated by an explicit
momentum suppression in the WIMP response function($\Q_{2}^{(7)}$,
$\Q_{3}^{(7)}$, $\Q_{4}^{(7)}$) or the steepness of the iodine form
factor can be mitigated by tuning the $r$ parameter ($\Q_{1,q}^{(6)}$,
$\Q_{2,q}^{(6)}$, $\Q_{5,q}^{(7)}$, $\Q_{6,q}^{(7)}$), or both
($\Q_{4,q}^{(6)}$,$\Q_{7,q}^{(7)}$, $\Q_{8,q}^{(7)}$ and
$\Q_{10,q}^{(7)}$).  Such mechanism is not altered by the effect of
the meson poles arising in the axial current of $\Q^{(6)}_{4,q}$, the
pseudoscalar current of $\Q^{(7)}_{7,q}$ and $\Q^{(7)}_{8,q}$ and the
CP--odd gluonic current of $\Q^{(7)}_{3,q}$ and $\Q^{(7)}_{4,q}$,
since they always induce a mild momentum dependence in the scattering
amplitude.

For all the minima the corresponding predicted number of events
exceeds by more than three orders of magnitude the upper bounds from
XENON1T and/or PICO--60.

\section*{Acknowledgements}
This research was supported through the Basic Science Research Program
of the National Research Foundation of Korea (NRF) funded by the
Ministry of Education, grant number 2016R1D1A1A09917964 and by the
Ministry of Science and ICT, grant number 2019R1F1A1052231.

\end{document}